\documentclass[12pt]{article}
\usepackage{amssymb,amsmath,epsfig}

\begin{document}
\begin{center}
{\Large\bf Gravitational Collapse and Expansion of Charged
Anisotropic Cylindrical Source\\}
\medskip

T. Mahmood $^{(a)}$ \footnote{e-mail: tahir12b@yahoo.com}, S. M.
Shah $^{(a)}$ \footnote{e-mail: syedmunawarshah71@hotmail.com} and
G.
Abbas$^{(b)}$\footnote{e-mail: ghulamabbas@ciitsahiwal.edu.pk}\\

$^a$\textit{Department of Mathematics, The Islamia University,\\
Bahawalpur-63100, Pakistan.}\\
$^b$\textit{{Department of Mathematics, COMSATS Institute of
\\Information
Technology Sahiwal-57000, Pakistan}}.\\

%\author{T. Mehmood $^a$ \thanks{tahir12b@hotmail.com}, S. M. Shah $^a$ \thanks{syedmunawarshah71@hotmail.com} and G. Abbas $^{a,b}$ \thanks{ghulamabbas@ciitsahiwal.edu.pk}\\
%%$^a$Department of Mathematics, The Islamia University,\\
%%Bahawalpur-63100, Pakistan.\\
%%$^b$ Department of Mathematics, GOVT. Post Graduate,\\ Collage Baghdad Road, Bahawalpur-63100,\\
%%Bahawalpur-63100, Pakistan.\\
%%$^c$ Department of Mathematis, COMSATS Institute of\\ Information
%%Technology Sahiwal-57000, Pakistan
\end{center}
\begin{abstract}
In this paper, we have discussed the gravitational collapse and
expansion of charged anisotropic cylindrically symmetric gravitating
source. To this end, the generating solutions of Einstein-Maxwell
field equations for the given source and geometry have been
evaluated. We found the auxiliary solution of the filed equations,
this solution involves a single function  which generates two kinds
of anisotropic solutions. Every solution can be expressed in terms
of arbitrary function of time that has been chosen arbitrarily to
fit the various astrophysical time profiles. The existing solutions
predict gravitational collapse and expansion depending on the choice
of initial data. Instead of base to base collapse, in the present
case, wall to wall collapse of the cylindrical source has been
investigated. We have found that the electromagnetic field is
responsible for the enhancement of anisotropy in collapsing system.
\end{abstract}
{\bf Keywords:} Cylindrical Symmetry; Gravitational Collapse;
Electromagnetic Field.\\
{\bf PACS:} 04.40.-b; 04.20.-q; 04.40.Dg.

\section{Introduction}
The study of gravitational collapse started from the pioneer work of
Chandrasekhar (1936). Later on, Oppenheimer and Snyder
(19390investigated that the spherically symmetric homogenous dust
collapse leads to the formation of black hole (BH). Initially, it
was argued that the homogeneity and spherical symmetry of the
collapsing model are responsible for the formation of BH. However,
it was found (Lemaitre 1933) that inhomogeneous dust collapse would
end as shell crossing and shell focusing singularities. After many
years, it was proved (Joshi 1997) that the shell crossing
singularity is naked, while shell focusing singularity might
represent BH, depending on the choice of initial data of collapse.
Hence, it was concluded that the homogeneity of the collapsing model
is not a sufficient condition for the formation of BH. In order to
generalize the collapsing matter, it becomes necessary to study the
collapse of more realistic matter with non-vanishing pressure.

Misner and Sharp (1964) studied perfect fluid collapse and found BH
as the end state of gravitational collapse. Also, Herrera and Santos
(1997) explored the properties of anisotropic self-gravitating
spheres and discussed their stability using the perturbation method.
Herrera and his collaborators (Herrera et al.2008a, Herrera et al.
2008b, Herrera at al. 1989, Herrera et al. 2009, Herrera et al.
2010, Herrera et al. 2012) have discussed the stability and
applications of anisotropic solutions to stellar collapse. Recently,
Glass (2013) has formulated a generating solution of anisotropic
spherically symmetric solutions which reveal either expansion and
collapse depending on the choice of time profile of the solutions.
This work has been extended for plane symmetric anisotropic source
and charged anisotropic sphere by Abbas (2014a, 2014b). The present
paper is the cylindrical version of the these papers.

The current observational evidences of gravitational waves (through
the detectors LIGO (Abramovici et al. 1992) and GEO (L$\ddot{u}$ck
and GEO 600 Team 1997) have increased the interest to study the
gravitational collapse in cylindrically symmetric systems. The
spherical systems are simple and do not provide non-trivial examples
in the generic gravitational collapse. Therefore, the study of
cylindrical collapse is much important as compared to spherical
systems. Some numerical studies (Piran 1978) provide the generation
of gravitational waves from cylindrical collapse. These results have
been extended analytically (Nakao and Morisawa 2004) to study
gravitational waves during cylindrical gravitational collapse.
Sharif and Ahmad (2007) generalized this work for two perfect fluid
cylindrical collapse and discussed the generation of gravitational
waves. Di Prisco et al.(2009) studied the shearfree gravitational
collapse of the anisotropic fluid in the cylindrically symmetric
spacetime.

The applications of the electromagnetic field in astronomy and
astrophysics is an active research domain. A lot of work has been
devoted to discuss the collective effects of electromagnetic and
gravitational fields. Till now, there is a little progress about the
effects of electromagnetic field on gravitational collapse of stars.
Thorne (1965) studied cylindrically symmetric gravitational collapse
with magnetic field and concluded that magnetic field can prevent
the collapse of cylinder before singularity formation. Ardvan and
Partovi (1977) investigated dust solution of the field equations
with electromagnetic field and found that the electrostatic force is
balanced by gravitational force during collapse of charged dust. The
effects of electromagnetic field on structure scalars and dynamics
of self-gravitating objects have been explored by Herrera and his
collaborators (Herrera et al. 2011, Diprisco et al. 2007). Sharif
and his collaborators (Sharif and Bhatti 2012a, Sharif and Bhatti
2012b, Sharif and Bhatti 2013a, Sharif and Bhatti 2013b, Sharif and
Bhatti 2014, Sharif and Yousaf 2012, Sharif and Kausar 2011) have
extended this work for cylindrical and plane symmetries in GR as
well as in $f(R)$ gravity with electromagnetic field.

This paper is organized as follows: In section \textbf{2}, charged
anisotropic cylindrical source and Einstein-Maxwell equations have
presented. Section \textbf{3} is devoted to the generating solutions
which represent gravitational collapse and expansion of the
self-gravitating charged cylinder. We summaries the results of the
paper in the last section.

\section{Matter Distribution and Field Equations}

This section deals with the interior matter distribution and
corresponding Einstein-Maxwell's equations. The non-static spacetime
with cylindrical symmetry in the interior region of a star is given
by (Sharif and Bhatti 2015)
\begin{equation}\label{1}
ds^2=-A^2dt^{2}+B^2dr^{2}+C^2{d\theta^{2}}+ dz^2,
\end{equation}
where $-\infty{<}t{<}\infty,~0\leq{r}<\infty,~
0\leq{\theta}\leq{2\pi},~-\infty<{z}<{\infty}$ are the restrictions
on the coordinates of cylinder. In the interior region of
cylindrically symmetric star, we have considered the charged
anisotropic fluid for which energy-momentum is given by (Di Prisco
et al. 2009)
\begin{eqnarray}\nonumber
T_{\alpha\beta}&=&(\mu+p_r)v_{\alpha}v_{\beta}
-(p_r-p_z)s_{\alpha}s_{\beta}-(p_r-p_\theta)k_{\alpha}k_{\beta}
\\\label{2}
&+&p_{r}g_{\alpha\beta}
+\frac{1}{4\pi}\left(F^{\gamma}_{\alpha}F_{\beta\gamma}-
\frac{1}{4}F^{\gamma\delta}F_{\gamma\delta}g_{\alpha\beta}\right),
\end{eqnarray}
where $\mu,~p_{r},~p_{\theta}$,and $p_{z}$ are the energy density,
pressures in $r,~\theta$ and $z$ directions, respectively. Further,
$v_{\alpha}$ is four-velocity and $s_{\alpha}$, $k_{\alpha}$ are
four-vectors. Also,
$F_{\alpha\beta}=-\phi_{\alpha,\beta}+\phi_{\beta,\alpha}$ is the
Maxwell field tensor with four-potential $\phi_\alpha$. Moreover,
$s_{\alpha}$ and $k_{\alpha}$ are the unit four-vectors which
satisfy the following relations
\begin{equation*}
s^{\alpha}s_{\alpha}=k^{\alpha}k_{\alpha}=1,\quad
v^{\alpha}v_{\alpha}=-1,\quad
s^{\alpha}k_{\alpha}=v^{\alpha}k_{\alpha}=v^{\alpha}s_{\alpha}=0.
\end{equation*}
In comoving coordinate system, these quantities can be written as
\begin{equation}\label{3}
k_{\alpha}=C{\delta}^{2}_{\alpha},\quad
v_{\alpha}=-A\delta^{0}_{\alpha},\quad
s_{\alpha}={\delta}^{3}_{\alpha}.
\end{equation}

The Maxwell field's equations are
\begin{equation*}\label{4}
F^{\alpha\beta}_{~~;\beta}={4\pi}J^{\alpha},\quad
F_{[\alpha\beta;\gamma]}=0,
\end{equation*}
where $J_\alpha$ is the four-current. In comoving coordinates, the
charge inside the cylinder is at rest, so we can define the
four-potential and four-current as follows:
\begin{equation*}
\phi_{\alpha}={\phi}{\delta^{0}_{\alpha}},\quad
J^{\alpha}={\zeta}v^{\alpha},
\end{equation*}
where $\zeta(r,t)$ and $\phi(r,t)$ are charge density and scalar
potential, respectively. The expansion scalar is
\begin{equation}\label{4}
\Theta=\frac{1}{A}\left(\frac{\dot{B}}{B}
+\frac{\dot{C}}{C}\right).
\end{equation}
We define the dimensionless anisotropy as follows:
 \begin{equation}\label{4a}
\Delta a=\frac{p_r-p_\theta}{p_r}.
\end{equation}
The corresponding Einstein-Maxwell's equations have the following
form
\begin{eqnarray}\label{5}
\kappa\left({\mu}-\frac{\pi}{2}E^2\right)A^{2}&=&\frac{\dot{B}\dot{C}}{BC}
+\left(\frac{A}{B}\right)^2
\left(\frac{B'C'}{BC}-\frac{C''}{C}\right), \\\label{6} 0&=&\frac{\dot{C'}}{C}-\frac{\dot{B}C'}{BC}-\frac{\dot{C}A'}{CA},
\end{eqnarray}
\begin{eqnarray}\label{8} \kappa\left({p_r}+\frac{\pi}{2}E^2\right)B^{2}
&=&\frac{A'C'}{AC}+\left(\frac{B}{A}\right)^2
\left(-\frac{\ddot{C}}{C}+\frac{\dot{A}\dot{C}}{AC}\right),
\\\label{9} \kappa\left({p_{\theta}}-\frac{\pi}{2}E^2\right)
&=&\left(\frac{1}{AB}\right)
\left(\frac{\dot{A}\dot{B}}{A^2}-\frac{A'B'}{B^2}-\frac{\ddot{B}}{A}+\frac{A''}{B}
\right),\\\nonumber
\kappa\left({p_z}-\frac{\pi}{2}E^2\right)
&=&-\frac{\ddot{B}}{A^2B}
+\frac{A''}{AB^2}-\frac{\ddot{C}}{A^2C}-\frac{A'B'}{AB^3}
+\frac{\dot{A}}{A^3}\left(\frac{\dot{C}}{C}
+\frac{\dot{B}}{B}\right)\\\label{10}
&-&\frac{C'}{B^2C}\left(\frac{B'}{B}+\frac{A'}{A}\right)
-\frac{\dot{B}\dot{C}}{A^2BC} +\frac{C''}{B^2C},
\end{eqnarray}
where $E=\frac{s}{2\pi C}$ with $s(r)=4\pi\int^r_{0}{\zeta}{BC}dr$ is the total amount of
charge per unit length of the cylinder.

Thorne (1965) defined the mass function for cylindrical geometry in
the form of gravitational C-energy per unit length of the cylinder.
The specific energy $m=\tilde{E}l$ ($l$ is the length of cylinder,
i.e., $g_{zz}$)  of cylindrical geometry (\ref{1}) in the presence
of electric charge is given by (Sharif and Bhatti 2015)
\begin{equation}\label{11}
m(t,r)=sC+\frac{1}{8}\left[1-\left(\frac{C'}{B}\right)^2+\left(\frac{\dot{C}}{A}\right)^2
\right].
\end{equation}
The auxiliary solution of Eq.(\ref{6}) is
\begin{equation}\label{12}
A=\frac{\dot{C}}{C^{\alpha}}, \quad B=C^{\alpha},
\end{equation}
where $\alpha$ is arbitrary constant. Now using Eq.(\ref{12}) in Eq.(\ref{4}), we get the following form of expansion scalar
\begin{equation}\label{13}
\Theta=(1+\alpha)C^{\alpha-1}.
\end{equation}
 For $\alpha>-1$ and $\alpha<-1$, we obtain expanding and collapsing solutions respectively. Using Eq.(\ref{12}), in Eqs.(\ref{5})-(\ref{10}), we get the following form of Einstein-Maxwell's Equations:
\begin{eqnarray} \label{14}
 8\pi\left(\mu-\frac{s^2}{8 \pi C^2}\right)&=&\alpha C^{2(\alpha-1)}+C^{-2 \alpha} \left(\frac{\alpha C' C''}{C^2}-\frac{C''}{C}\right),\\\label{15}
 8\pi \left(p_r+\frac{s^2}{8 \pi C^2}\right)&=&\alpha C^{2(2\alpha-1)}+\frac{C'\dot{C}'}{\dot{C}C^{(2\alpha+1)}}-\frac{\alpha C'}{C^{(2\alpha+2)}},\\\nonumber
 8\pi \left(p_\theta+\frac{s^2}{8 \pi C^2}\right)&=&\alpha C^{2(\alpha-1)}\left(\frac{C\ddot{C}}{{\dot{C}}^2}-\alpha\right)-\left(\frac{ \dot{C}'C-\alpha\dot{C}C'}{C^{2(\alpha+1)}}\right)C'\\\nonumber&-&\alpha C^{\alpha}\left((\alpha-1)(\frac{\dot{C}}{C})^2 +\frac{\ddot{C}C^{\alpha}}{C\dot{C}}\right)+\left(\frac{(1-\alpha)(C''\dot{C}+\dot{C'}C')}{C^{2\alpha+1}}\right)\\\label{16} &-&\left(\frac{(\alpha+1)C'(C \dot{C'}-\alpha{\dot{C}}C')}{{C^{(\alpha+2)}}}\right),
\end{eqnarray}
 \begin{eqnarray}\nonumber
 8\pi \left(p_z+\frac{s^2}{8 \pi C^2}\right)&=&\left(\frac{\alpha(\alpha-1)}{C^2}+\frac{\alpha\ddot{C}}{C\dot{C}^2}\right)C^{2\alpha}+\frac{(1-\alpha)(C''\dot{C}+C'\dot{C'})}
 {\dot{C}C^{3\alpha+1}}\\\nonumber&-&\frac{(\alpha+1)\left(CC'\dot{C'}-\alpha\dot{C}C'^2\right)}{C^{2(\alpha+1)}\dot{C}}-\frac{\ddot{C}C^{2\alpha-1}}{{\dot{C}}^2}-\frac{\alpha \dot{C'}C'C}{C^{2{\alpha+1}}\dot{C}}\\\nonumber&+&\frac{{\alpha}^2{C'}^2}{C^{2(\alpha+1)}}+\frac{(\alpha+1){C}^{2(2\alpha-1)}\left(C\ddot{C}-\alpha {\dot{C}}^2\right)}{\dot{C}^2}+\frac{C''-\alpha
 C^{3(\alpha-1)}}{C^{2\alpha+1}}\\\label{17}&-&\frac{CC'\dot{C'}}{\dot{C}C^{2(\alpha+1)}}.
 \end{eqnarray}
For specific values of $C(r,t)$ and $\alpha$, we can find
anisotropic configuration. In this case mass function along with
electromagnetic field given in Eq.(\ref{11}) takes the following
form:
\begin{equation}\label{18}
{8m}-{8sC}-1=C^{2\alpha}-\frac{C'^2}{C^{2\alpha}}
\end{equation}
If $C'=C^{2\alpha}$ then above equation gives
\begin{equation}\label{19}
C=\frac{1}{s}\left(m-\frac{1}{8}\right)
\end{equation}
where $m>\frac{1}{8}$. This implies that gravitational collapse
leads to the formation of a trapping surface at
$C=\frac{1}{s}\left(m-\frac{1}{8}\right)$. Also, the integration of
trapping condition $C'=C^{2\alpha}$ yields
\begin{equation}\label{20}
C^{(1-2\alpha)}=(1-2\alpha)r+h(t),
\end{equation}
where $h(t)$ is an arbitrary function.

 \begin{figure}
\center\epsfig{file=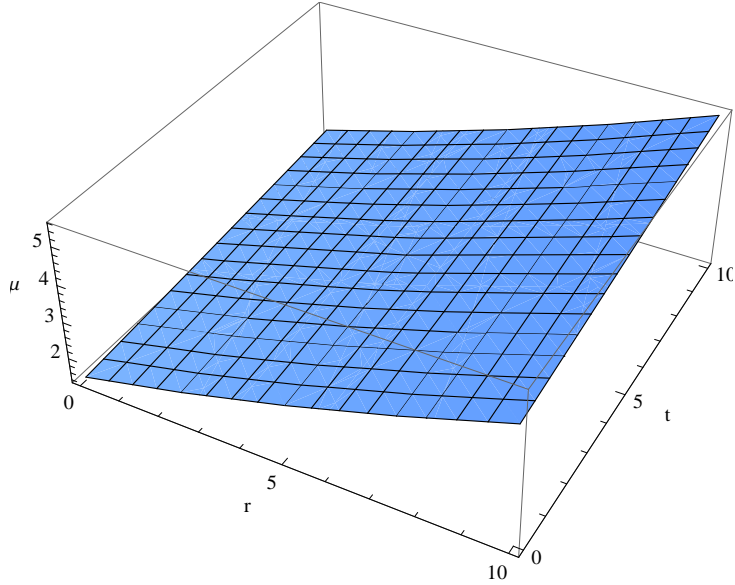, width=0.7\linewidth} \caption{Density
variation for $s=2$ and $h_1(t)=1+t.$}
\end{figure}
 \begin{figure}
\center\epsfig{file=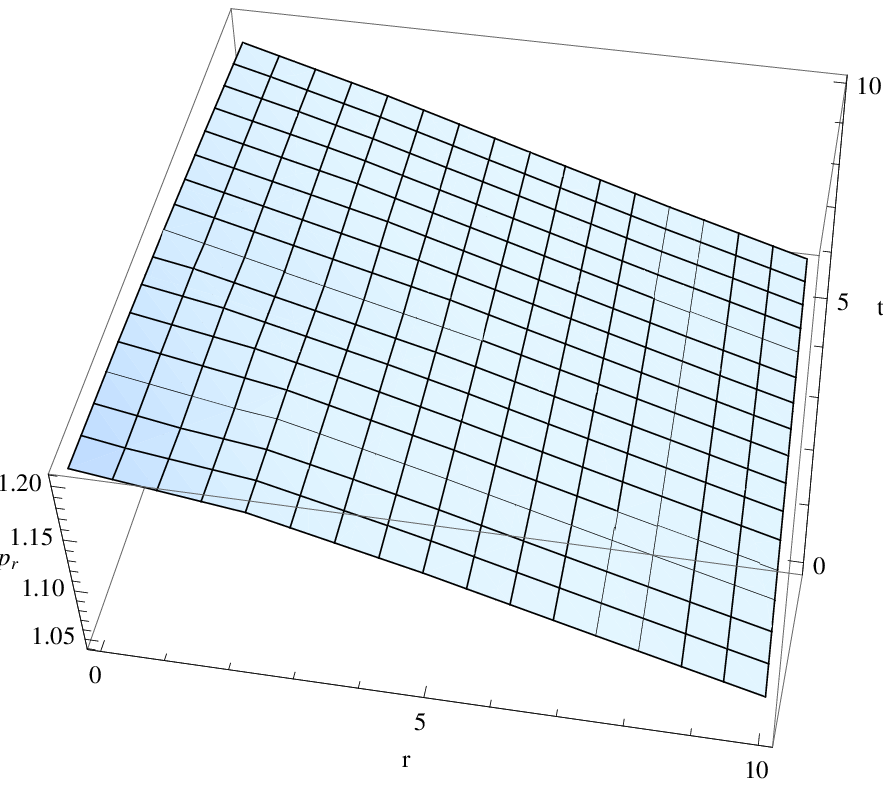, width=0.7\linewidth} \caption{Radial
pressure variation for $s=2$ and  $h_1(t)=1+t.$}
\end{figure}
\begin{figure}
\center\epsfig{file=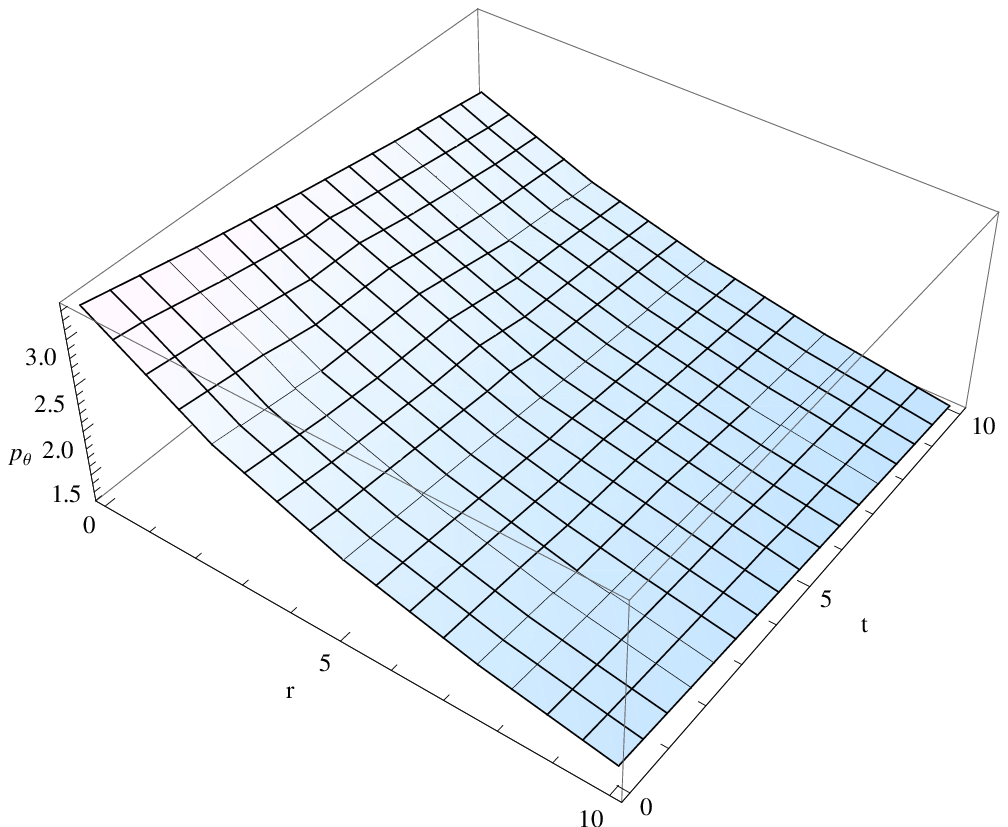, width=0.7\linewidth} \caption{Transverse
pressure variation for $s=2$ and  $h_1(t)=1+t.$}
\end{figure}
\begin{figure}
\center\epsfig{file=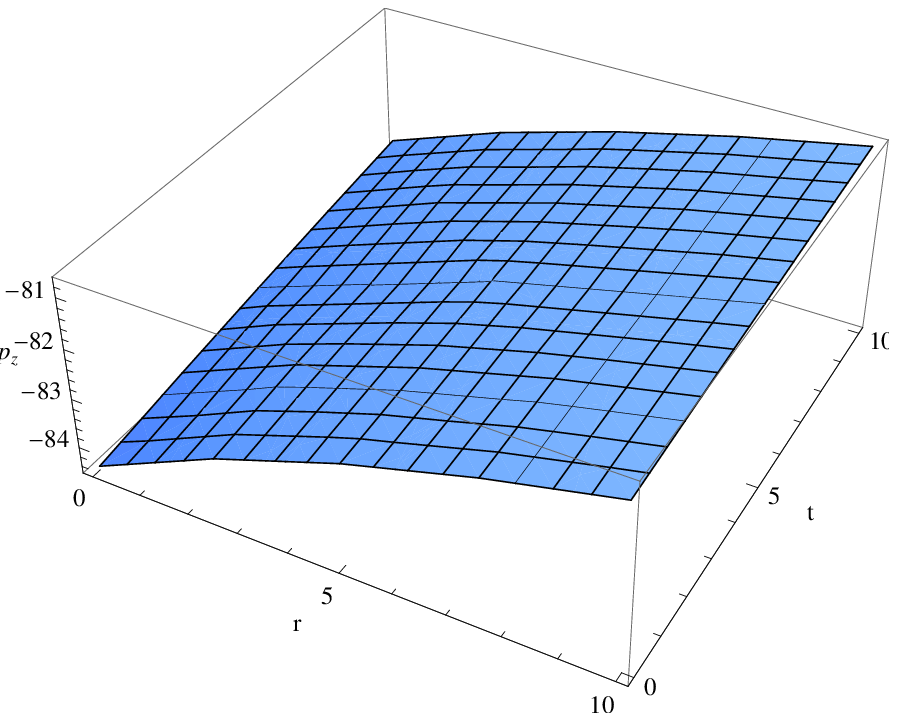, width=0.7\linewidth}
\caption{Longitudinal pressure variation for $s=2$ and
$h_1(t)=1+t.$}
\end{figure}
\begin{figure}
\center\epsfig{file=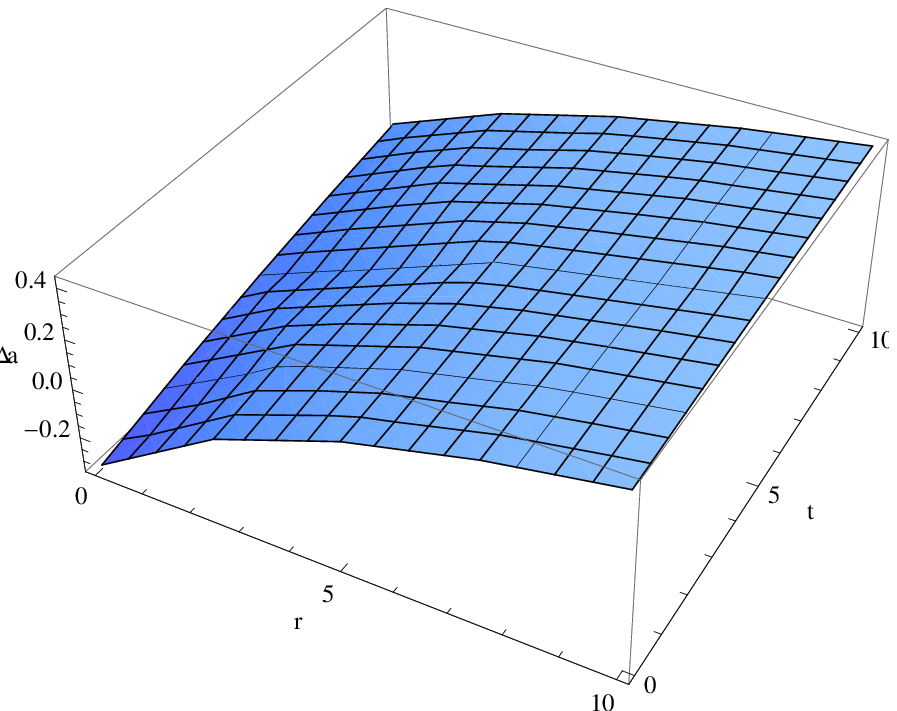, width=0.7\linewidth} \caption{
Dimensionless anisotropic parameter variation for $s=2$ and
 $h_1(t)=1+t.$ }
\end{figure}
\section{Generating solution}
For negative and positive values of $\alpha$ , we have collapsing
and expanding solutions, respectively as follows:
\subsection{Gravitational Collapse with
$\alpha=-\frac{3}{2}$} For collapse, expansion scalar will be
negative, from Eq.(\ref{11}),$\Theta<0$ when $\alpha>-1$. We assume
that $\alpha=-\frac{3}{2}$ and the condition $C'=C^{2\alpha}$, leads
to $C'=C^-3$, the integration of this equation yields,
\begin{equation}\label{21}
C_{trap}=(4r+h_1(t))^\frac{1}{4},
\end{equation}
where $h_1(t)$ is arbitrary function of time. For
$\alpha=-\frac{3}{2}$, Eqs.(\ref{14})-(\ref{17}) are given by
\begin{eqnarray} \label{21}
8\pi\mu&=&-C^3\left(\frac{3C'C''+CC''}{2C^2}\right)-\frac{3C^{-3}-s^2}{2}\\\label{22}8\pi
p_r&=&\frac{C\left(3C'\dot{C}+CC'\dot{C'}\right)}{2\dot{C}}
-\frac{3C^{-5}-s^2}{2C^2}\\\nonumber8\pi
p_\theta&=&\frac{C\left(5CC''\dot{C}+3CC'\dot{C'}-3{C'}^2\dot{C}\right)}{2}+\frac{2C^4C'\dot{C'}+3C^3C'^2\dot{C}}{4C^\frac{7}{2}}\\\label{24}
&-&\frac{15{\dot{c}}^2}{4C^\frac{7}{2}}+\frac{6C\dot{C}\ddot{C}-6\ddot{C}C+9{\dot{C}}^2}{4C^3{\dot{C}}^2}-\frac{s^2}{C^2}\\\nonumber
8\pi
p_z&=&\left(\frac{2C^5{\dot{C}}^2+3C^3C'\dot{C}+6C^3C'\dot{C'}\dot{C}+9C^3C'^2\dot{C}+6C^3{\dot{C}}^2+2C^4C'\dot{C'}}{4C^2\dot{C}}\right)\\\nonumber
&+&\frac{6C^4{\dot{C}}^+4C^4C''\dot{C}+4s^2\dot{C}}{4C^2\dot{C}}+{C^\frac{1}{5}}\left(\frac{12\ddot{C}-15{\dot{C}}^2}{4{\dot{C}}^2}\right)-
\frac{\ddot{C}}{C^4{\dot{C}}^2}-\frac{C\ddot{C}+6{C}^2}{4C^8{\dot{C}}^2}\\\label{26}&+&C^{-\frac{7}{2}}\left(\frac{3}{2}+\frac{5}{2}
(\frac{C''}{C}+\frac{C'{\dot{C'}}^2}{C\dot{C}})\right).
\end{eqnarray}
Using Eq.(\ref{21}) in the above equations, we get the following
form of field equations:
\begin{eqnarray} \label{26}
8\pi \mu &=&-\frac{3}{2 (4 r+h_1)^{5/4}}+\frac{s^2}{\sqrt{4
r+h_1}}+\left(\frac{9+6(4 r+h_1)}{2 (4
r+h_1)^{\frac{9}{4}}}\right),\\\label{27}8\pi
P_r&=&-\frac{s^2}{\sqrt{4 r+h_1}}+\frac{3}{2} \left(-\frac{1}{(4
r+h_1)^2}-\frac{2}{(4 r+h_1)^{5/4}}+\frac{1}{\sqrt{4
r+h_1}}\right),\\\nonumber 8\pi p_{\theta }&=&\left(\frac{3
\left(144 (4 r+h_1)^{9/8} \dot{h_1}^2-4 \left(1+24 (4
r+h_1)^{3/8}\right) \dot{h_1}^3\right.}{64 (4 r+h_1)^{19/8}
\dot{h_1}^2}\right)
\\\label{27a} &-&\left(\frac{\left.5 h_1'^4-128 (4 r+h_1)^{17/8}
\ddot{h_1}+32 (4 r+h_1)^{11/8} \dot{h_1} \ddot{h_1}\right)}{64 (4
r+h_1)^{19/8} \dot{h_1}^2}\right),
\end{eqnarray}
\begin{eqnarray}\nonumber 8\pi p_z&=&\frac{4 s^2}{\sqrt{4
r+h_1}}+\frac{3 }{4}\left(-\frac{20}{(4 r+h_1)^{23/8}}+\frac{1}{(4
r+h_1)^2}-\frac{5}{(4 r+h_1)^{5/4}}\right)\\\nonumber
&+&\frac{3}{2(4 r+h_1)^{7/8}}-\frac{12}{(4 r+h_1)^{1/5}}-5 (4
r+h_1)^{1/20}\\\nonumber &+&\frac{1}{8} \left(1-\frac{3}{(4
r+h_1^2}+\frac{3}{\sqrt{4 r+h}}\right) \dot{h_1}+\frac{15
\dot{h_1}^2}{64 (4 r+h_1)^{5/4}}\\\label{28}&+&\frac{2
\left(-\frac{1}{4 r+h_1}-\frac{2}{(4 r+h_1)^{1/4}}+6 (4
r+h_1)^{4/5}\right)\dot{h_1}}{\dot{h_1}^2}.
\end{eqnarray}
The dimensionless measure of anisotropy is given by the following equation:
\begin{eqnarray}\nonumber
\text{$\Delta $a}&=&\left[\frac{1 \left(-\frac{6}{(4
r+h_1^2}-\frac{39}{(4 r+h_1^{5/4}}+\frac{6} {\sqrt{4 r+h_1}}-\frac{4
s^2}{\sqrt{4 r+h_1}}\right)}{4 \left(-\frac{s^2}{\sqrt{4
r+h_1}}+\frac{3}{2} \left(-\frac{1}{(4 r+h_1)^2}-\frac{2}{(4
r+h_1)^{5/4}}+\frac{1}{\sqrt{4 r+h_1}}\right)\right)}\right]
\\\nonumber &+& \left[\frac{\left[\left(12+96 (4 r+h_1)^{3/8}\right)\dot{ h_1}^3+5 \dot{h_1}^4\right]}{64 \left(-\frac{s^2}
{\sqrt{4 r+h_1}}+\frac{3}{2} \left(-\frac{1}{(4
r+h_1)^2}-\frac{2}{(4 r+h_1)^{5/4}}+\frac{1}{\sqrt{4
r+h_1}}\right)\right.} \right]\\\label{a11} &+& \left[\frac{128 (4
r+h_1)^{17/8} \dot{h_1}-32 (4 r+h_1)^{11/8} \dot{h_1} \ddot{h_1}}{64
\left(-\frac{s^2}{\sqrt{4 r+h_1}} +\frac{3}{2} \left(-\frac{1}{(4
r+h_1)^2}-\frac{2}{(4 r+h_1)^{5/4}}+\frac{1}{\sqrt{4
r+h_1}}\right)\right)}\right].
\end{eqnarray}

For $\alpha=\frac{-5}{2}$, we obtain $\Theta<0$ and matter density
increases for the arbitrary choice charge $s$ and time profile
$h_1=1+t$. As density is increasing (see figure.\textbf{1}), so
cylinder goes on collapsing to a point. In this case, the length of
cylinder is constant, i.e., $g_{zz}=1$, therefore base to base
collapse is impossible, there is only possibility of wall to wall
collapse. The anisotropic parameter $\Delta a$ changes its sign from
negative to positive. The anisotropy will be directed outward when
$p_\theta>p_r$, this implies that $\Delta a>0$ and directed inward
when $p_\theta<p_r$ implying $\Delta a<0$. In this case $\Delta
a>0$, for larger value of $r$ as shown in figures \textbf{5}. This
implies that anisotropic force allows the construction of more
massive star while $\Delta a<0$ near the center, so there exist an
attractive force. When we talk about an external (electromagnetic)
field in gravitational, then there is an external force which may
distort the generic properties of spacetime effectively. So, in the
present case electromagnetic field enhances the anisotropy and the
homogeneity of collapsing star.

\subsection{Expansion with $\alpha=\frac{3}{2}$}
We know that for expansion, the expansion scalar will be positive,
from Eq.(\ref{11}), $\Theta<0$,  when $\alpha>0$. In this case
assume that $C=(r^2+r_0^2)^{-1}+h_2(t)$, where $h_2(t)$ and $r_0$
are arbitrary function and constant respectively. For
$\alpha=\frac{3}{2}$, Eqs.(\ref{14})-(\ref{17}) take the following
form:
\begin{eqnarray}
8\pi \mu&=&
\frac{s^2}{C^2}+\frac{3C}{2}+C^{-3}\left(\frac{3C'C''}{2C^2}-\frac{C''}{C}\right),\\\label{33}
8\pi
p_r&=&-\frac{s^2}{C^2}+\frac{3C^4}{2}+\frac{C'\dot{C'}}{C^4\dot{C}}-\frac{3C'}{2C^5},
\\\nonumber 8\pi p_{\theta} &=& -\frac{s^2}{C^2}+\frac{3C'}{2}\left(\frac{C\ddot{C}}{{\dot{C}}^2}-\frac{3}{2}\right)-\left(\frac{\dot{C}C'-\frac{3}{2}\dot{C}C'}{C^5}\right)C'-\frac{3}
{2}\left(\frac{{\dot{C}}^3+2C\ddot{C}}
{2C^{\frac{3}{2}}}\right)\\\label{34}&+&\frac{\left(C''C'+\dot{C'}C'\right)-5C^{\frac{1}{2}}C'\left(C\dot{C'}-\frac{3}{2}C'\dot{C}\right)}{2C^4},
\\\nonumber
8\pi p_z
&=&\frac{s^2}{C^2}-\frac{3}{2}\left(\frac{1}{2C}+\frac{\ddot{C}}{{\dot{C}}^2}\right)C^2-\frac{\left(C''\dot{C}+C'\dot{C'}\right)-5C'C''\left
(CC''-\frac{3}{2}C'\dot{C}\right)}{2C^{\frac{11}{2}}\dot{C}}\\\nonumber
&-&\frac{C^2\ddot{C}}{{\dot{C}}^2}-\left(\frac{6CC'\dot{C'}-9\dot{C}{C'}^2}{4C^5\dot{C}}\right)+\frac{C^5\ddot{C}}{2{\dot{C}}^2}-\frac{15C^4}{4}-
C^{-5}\left(\frac{3C'^2+2C\dot{C'}-3C'\dot{C}}{2}\right)
\\\label{35}&-&\frac{3C^{\frac{-1}{2}}}{2}+\frac{C''}{C^4}.\label{35}
\end{eqnarray}
If $F(t,r)=1+h_2(t)(r^2+r_0^2)$ and $C=\frac{F}{r^2+r_0^2}$ then
Eqs.(\ref{33})-(\ref{35}) become:
\begin{eqnarray}
{8\pi}
\mu&=&\left(\frac{s^2(F-1)^2}{h_2(t)F^2}\right)+\frac{3h_2(t)F}{8\pi(F-1)}-\frac{2(r_0^2-3r^2)(F^2-F-3r)}{8\pi
h_2(t)F^5}\\\label{37} {8\pi}
p_r&=&\left(-{\left(\frac{s(F-1)}{h_2(t)F}\right)^2}\right)+\frac{3}{16\pi}\left(\frac{F
h_2(t)}{F-1}\right)^4+\frac{3r}{8\pi
F^5}\left(\frac{F-1}{h_2(t)}\right)^3\\\nonumber {8\pi}
p_{\theta}&=&\frac{1}{2}\left(\left(-\frac{2s(r^2+r^2_0)}{F}\right)^2+
\frac{2r(6r^2-2r^2_0+15r(r^2+r^2_0)\sqrt{F})\dot{h_2(t)}}{(r^2+r^2_0)^\frac{3}{2}F^4}\right)
\\\nonumber &+&\frac{4r^2(r^2+r^2_0)\dot{h_2(t)}}{F^5}+\frac{3}{2}\left((\frac{F}{(r^2+r^2_0)})(-\frac{3}{2}+\frac{F \ddot{h_2}}{{h_2}^2(r^2+r^2_0)})\right)\\\label{38}&-&\frac{3}{2}\left(\left(\frac{F}{(r^2+r^2_0)}\right)^\frac{3}{2}\right)
\left(\frac{{\dot{h_2(t)}}^2(r^2+r^2_0)^2}{2F^2}+\frac{\sqrt{F}\ddot{h_2}}{\sqrt{(r^2+r^2_0)}\dot{h_2}}\right)
\end{eqnarray}
\begin{eqnarray}\nonumber {8\pi} p_z&=&\frac{1}{2}\left[\left(\frac{s(r^2+r^2_0)}{F}\right)^2-\frac{3\sqrt{(r^2+r^2_0)}}{\sqrt{F}}+\frac{18r^2(r^2+r^2_0)}{F^5}
-\frac{4(r^2+r^2_0)(r^2_0-3r^2)}{F^4}\right]\\\nonumber
&-&\frac{6r(r^2+r^2_0)(2r+(r^2+r^2_0)^2)\dot{h_2}}{F^5}+\frac{5F^4(-\frac{3}{2}{\dot{h_2}}^2+(r^2+r^2_0)F
\ddot{h_2}}{(r^2+r^2_0)^5(\dot{h_2})^2}\\\nonumber
&-&\frac{2(-3r^2+r^2_0)\left(20r+20r(-3r^4+r^4_0-2r^2r^2_0)h_2+(r^2+r^2_0)^2(r^8+r^8_0)
\dot{h_2}\right)}{(r^2+r^2_0)^\frac{7}{2}F^\frac{11}{2}\dot{h_2}}\\\nonumber
&+&\frac{(r^2+r^2_0)^2(4r^6r^6_0+6r^4r^4_0+4r^2_0r^6-30r^2)
\dot{h_2}}{(r^2+r^2_0)^\frac{7}{2}F^\frac{11}{2}\dot{h_2}}-\frac{2F^2\ddot{h_2}}{(r^2+r^2_0)^2\dot{h_2}}\\\label{39}
&-&\frac{3F}{2(r^2+r^2_0)^2}\left(1+\frac{2F
\ddot{h_2}}{(r^2+r^2_0)^2\dot{h_2}}\right).
\end{eqnarray}
The dimensionless measure of anisotropy in this case takes the following form:
\begin{eqnarray}\nonumber
\Delta
a&=&\left[\frac{3F^4}{(r^2+r^2_0)^4}+\frac{6r(r^2+r^2_0)^3}{F^5}-\frac{4r^2(r^2+r^2_0)\dot{h_2}}{F^5}+
\left(\frac{3F^\frac{3}{2}}{(r^2+r^2_0)^\frac{3}{2}}\right)\right]\\\nonumber
&-&\left[\frac{2r(-2r^2_0+6r^2+15r\sqrt{(r^2+r^2_0)F}\dot{h_2})}{(r^2+r^2_0)F^4}
-(\frac{3F}{(r^2+r^2_0)})\left(\frac{-3}{2}+\frac{F\ddot{h_2}}{(r^2+r^2_0){\dot{h_(t)}}^2}\right)\right]\\
&\times&\left[\frac{(r^2+r^2_0){\dot{h_(t)}^2}}{2F^2}+\frac{\sqrt{F}\ddot{h_2}}{\sqrt{(r^2+r^2_0)}\dot{h_2}}\right]
\left[\frac{1}{-\frac{s^2((r^2+r^2_0)^2)}{F^2}+\frac{3}{2}\left(\frac{F^4}{(r^2+r^2_0)^4}\right)+\frac{3r(r^2+r^2_0)^3}{F^5}}\right].
\end{eqnarray}

 \begin{figure}
\center\epsfig{file=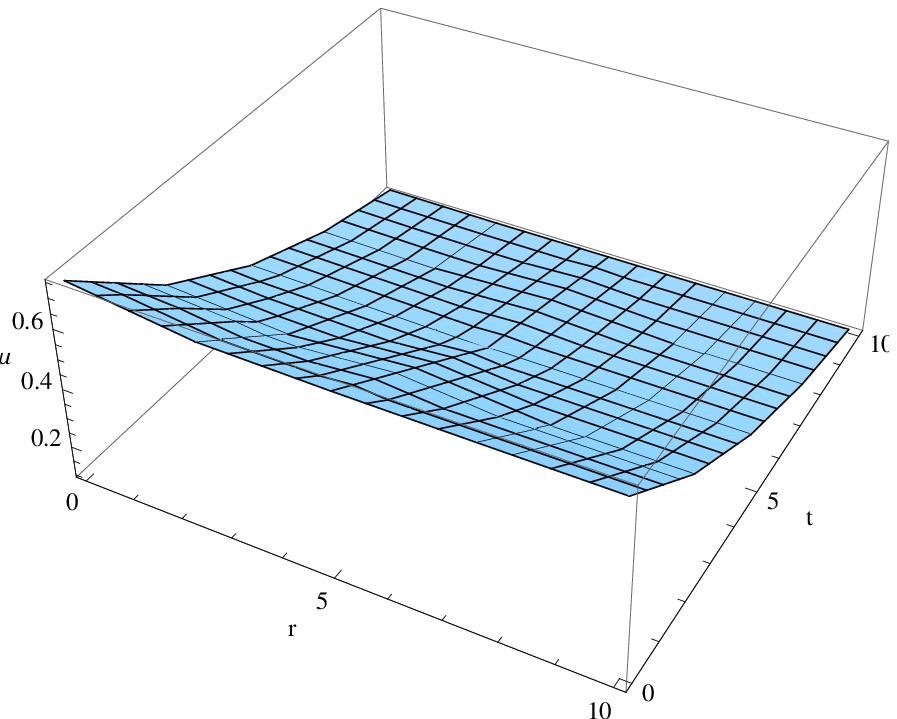, width=0.7\linewidth} \caption{Density
variation for $s=2$ and $h_2(t)=1+t.$}
\end{figure}
 \begin{figure}
\center\epsfig{file=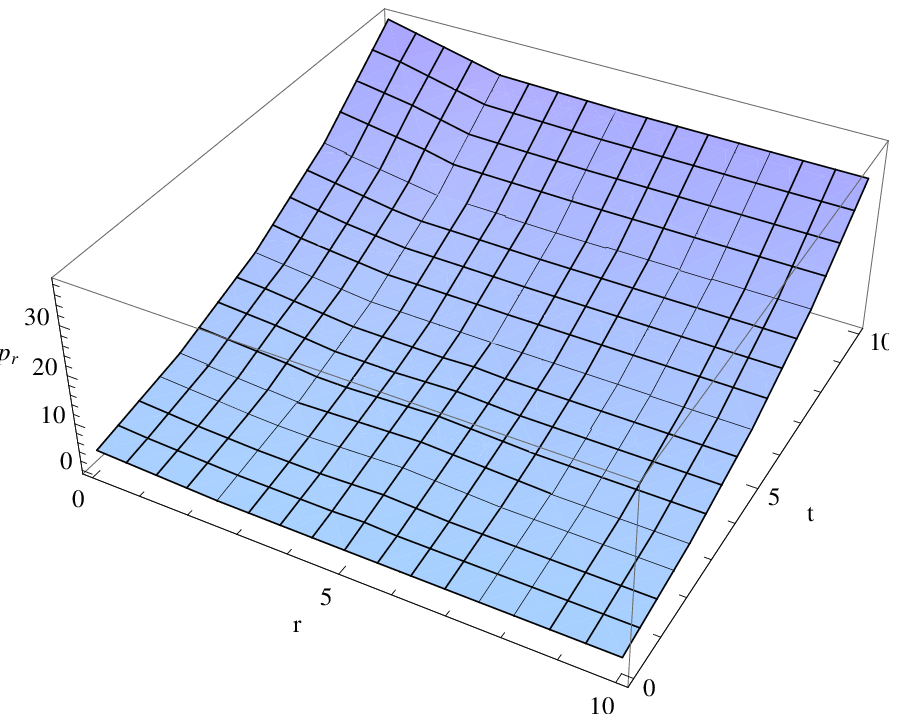, width=0.7\linewidth} \caption{Radial
pressure variation for $s=2$ and $h_2(t)=1+t.$}
\end{figure}
\begin{figure}
\center\epsfig{file=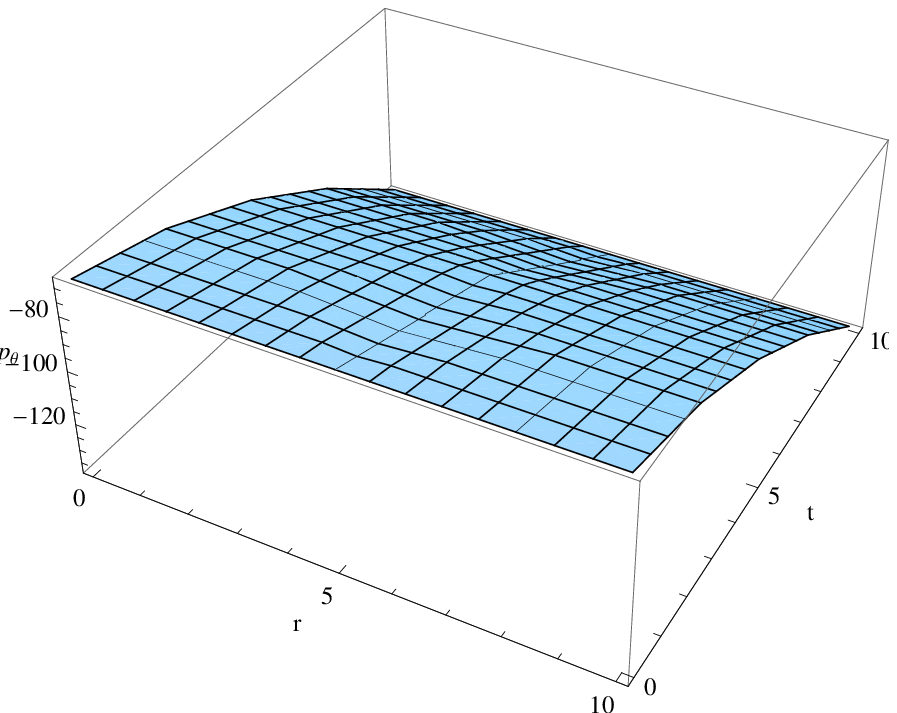, width=0.7\linewidth} \caption{Transverse
pressure variation for $s=2$ and $h_2(t)=1+t$.}
\end{figure}
\begin{figure}
\center\epsfig{file=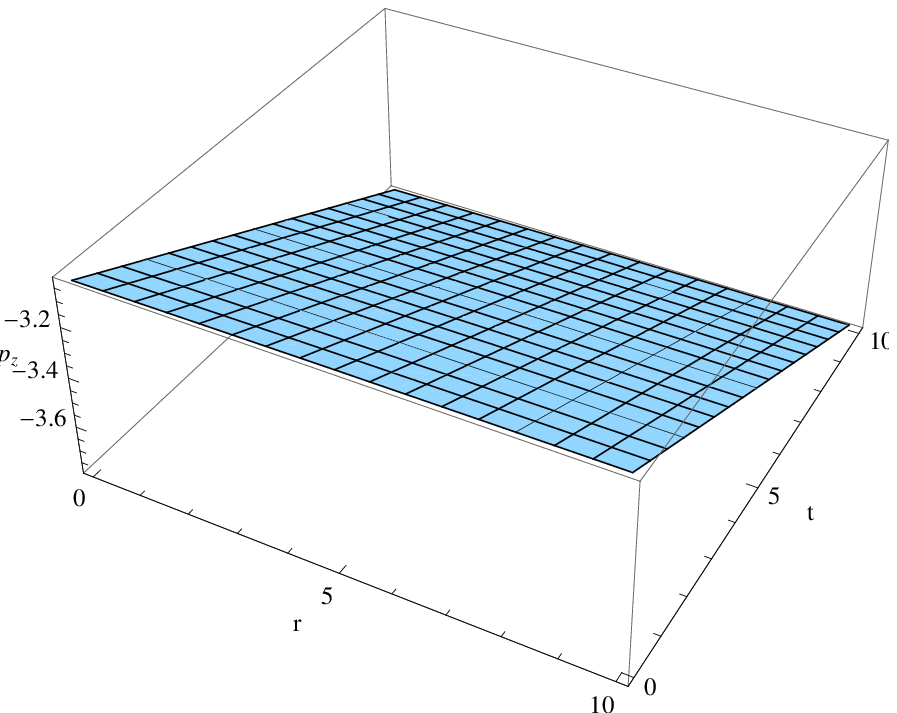, width=0.7\linewidth}
\caption{Longitudinal pressure variation for $s=2$ and
$h_2(t)=1+t.$}
\end{figure}
\begin{figure}
\center\epsfig{file=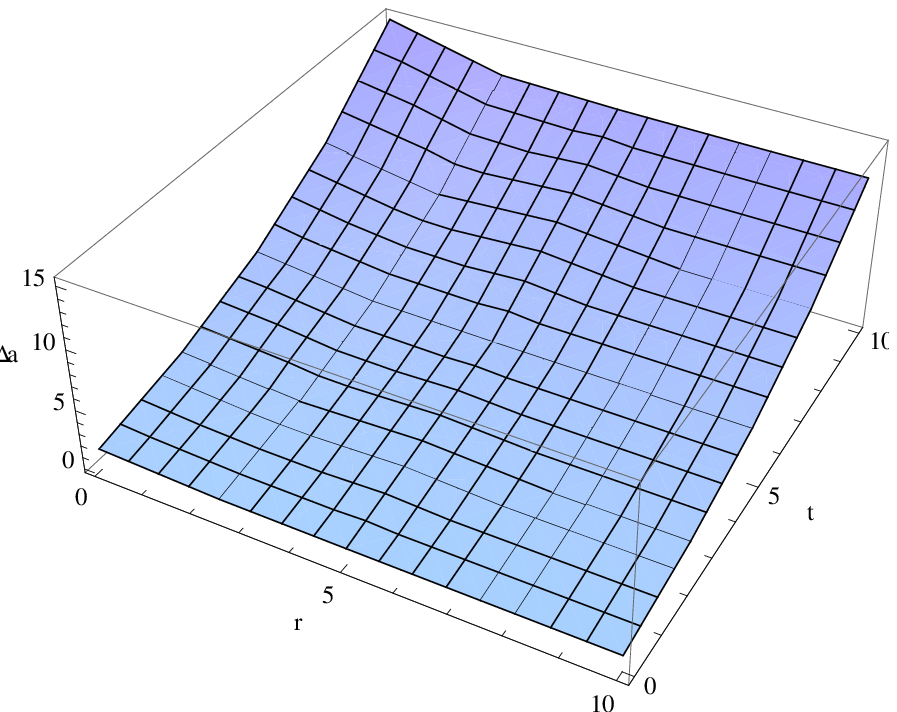, width=0.7\linewidth} \caption{
Dimensionless anisotropic parameter variation for $s=2$ and
$h_2(t)=1+t.$}
\end{figure}

For $\alpha=\frac{3}{2}$, we have $\Theta>0$ and matter density
decreases (see figure\textbf{ 6}) for the arbitrary choice of charge
and time profile $h_2=1+t$. In this case $p_r$ and $p_\theta$ are
positive and negative (see figure \textbf{7,8}), respectively. This
implies that there exist anisotropy due to opposite behaviour of
pressure components. From figure \textbf{10} $\Delta a>0$, there
exists repulsive force which causes the expansion of matter in this
case. The expansion process causes to separate the charges apart
from each other, this results to weak electromagnetic field
intensity.

\section{Concluding Remarks}

During the last few decades, there has been a growing interest to
study the relativistic anisotropic systems due to the existence of
such systems in astronomical objects. The exact solutions of
anisotropic sources are helpful to determine the anisotropy of the
universe during any era. The effects of anisotropy on the late-time
expansion of inhomogeneous universe have been studied by Barrow and
Maartens (1998). They remarked that the decrease in shear anisotropy
can be figure out by measuring the anisotropic pressure of the
cosmological model. Herrera and Santos (1997) pointed out that the
phase transition in anisotropic highly dense system would occur
during the gravitational collapse. Further, they concluded that such
system may transited to a pion condensed phase, where a soften
equation of state can provide enough exhausted energy.

This paper is aimed to study the generating solution of
Einstein-Maxwell field equations with anisotropic cylindrically
symmetric fluid. We have used the auxiliary solution of one field
equation to determine the solution of the remaining equations. The
application of assumed solution in expansion scalar, allows us to
determine the range of free constant $\alpha$, for which expansion
scalar $\Theta$ is positive or negative, leading to expansion and
collapse. The C-energy analogous to Misner-Sharp mass has been
calculated with the contribution. We have impose the condition
${C'}=C^{2\alpha}$, on the mass function which leads to the
existence of trapping horizon at
$C=\frac{1}{s}\left(m-\frac{1}{8}\right)$, provided $m>\frac{1}{8}$.
In this case curvature singularity is hidden at the center of
trapping horizon.

The expansion scalar $\Theta=(2+\alpha)C^{(1-\alpha)}$, becomes
$\Theta=0$, for $\alpha=-1$, $\Theta>0$, for $\alpha>-1,$
$\Theta<0$, for $\alpha<-1$, which corresponds to bouncing,
expansion and collapse, respectively. In other words $\Theta>0$ ,
$\alpha\in (0,\infty),$ and $\Theta<0$ for $\alpha\in (-\infty,0)$
which corresponds to expansion and collapse, respectively. For the
sake of simplicity, we have taken $\alpha=-\frac{3}{2}$ for
gravitational collapse and $\alpha=\frac{3}{2}$ for expansion,
explicitly. The full dynamics of the system has been discussed in
both cases. The matter density is increasing/dcreasing function with
arbitrary choice of charge parameter and time profiles. The
pressures $p_r,~~{p_\theta}$ and $p_z$ are different in both cases,
therefore the pressure anisotropy is non-vanishing in both cases.
This anisotropy is increasing function in both cases, it is due to
presence of electromagnetic field, as it produces an external
repulsive force to distort the geometry of the star.

We would like to mention this work with \textbf{charged plane
symmetric source} is in progress.
\section{Conflict of Interest} The authors declare that they have no
conflict of interest.

\end{document}